\documentclass{ws-p8-50x6-00}

\begin{document}

\title{Quasifree processes from nuclei: Meson photoproduction and electron scattering}

\author{L.J. Abu-Raddad${}^{1}$
\footnote{Electronic address: laith@rcnp.osaka-u.ac.jp} 
and 
J. Piekarewicz${}^{2}$
\footnote{Electronic address: jorgep@csit.fsu.edu}}
\address{${}^{1}$ Theory Group, Research Center for Nuclear Physics,\\
Osaka University, 10-1 Mihogaoka, Ibaraki City, Osaka 567-0047, Japan}

\address{${}^{2}$Department of Physics, \\
         Florida State University, 
         Tallahassee, FL 32306, USA}


\maketitle

\abstracts{We have developed a relativistic formalism for studying quasi-free
processes from nuclei. The formalism can be applied with ease to a
variety of processes and renders transparent analytical expressions
for all observables. We have applied it to kaon photoproduction and to
electron scattering. For the case of the kaon, we compute the recoil
polarization of the lambda-hyperon and the photon asymmetry. Our
results indicate that polarization observables are insensitive to
relativistic, nuclear target, and distortion effects. Yet, they are
sensitive to the reactive content, making them ideal tools for the
study of modifications to the elementary amplitude --- such as in the
production, propagation, and decay of nucleon resonances --- in the
nuclear medium. For the case of the electron, we have calculated the
spectral function of ${}^{4}$He. An observable is identified for the clean
and model-independent extraction of the spectral function.  Our
calculations provide baseline predictions for the recently measured,
but not yet fully analyzed, momentum distribution of ${}^{4}$He by the
$A1$-collaboration from Mainz. Our approach predicts momentum
distributions for ${}^{4}$He that rival some of the best non-relativistic
calculations to date.}

\section{Introduction}

%\subsection{Producing the Hard Copy}\label{subsec:prod}

Faced by an increasing demand for studying quasifree processes from
nuclei, we have developed a general fully relativistic
treatment for studying such
interactions~\cite{abpi2000,thesis,abpi2001}. The power of the theoretical approach employed here lies in its
simplicity. Analytic expressions for the response of a mean-field
ground state may be provided in the plane-wave limit. The added 
computational demands placed on such a formalism, relative to that 
from a free on-shell proton, are minimal. The formalism owes its
simplicity to an algebraic trick, first introduced by Gardner 
and Piekarewicz~\cite{gp94}, that enables one to define a ``bound'' 
(in direct analogy to the free) nucleon propagator. Indeed, the 
Dirac structure of the bound nucleon propagator is identical to 
that of the free Feynman propagator. As a consequence, the power
of Feynman's trace techniques may be employed throughout the 
formalism.

We have applied this formalism
to two kinds of processes: kaon photoproduction~\cite{abpi2000,thesis} and electron
scattering~\cite{abpi2001}. Further, there is a promising potential of applying it to many
processes being studied experimentally at various laboratories. 
We will give here a brief introduction to this formalism
and we will discuss some of the results of using it.

The investigation of the quasifree kaon photoproduction process is impelled by recent
experimental advances and the increasing
interest in the study of strangeness-production reactions from
nuclei. These reactions form our gate to the relatively unexplored
territory of hypernuclear physics. Moreover, these reactions
constitute the basis for studying novel physical phenomena, such as
the existence of a kaon condensate in the interior of neutron
stars\cite{kn86}.

As for electron scattering, the appeal of this reaction is due to the perceived 
sensitivity of the process to the nucleon momentum
distribution. Interest in this reaction has stimulated a tremendous
amount of experimental work at electron facilities such as NIKHEF, 
MIT/Bates, and Saclay, who have championed this effort for several 
decades. Our motivation for studying this process is twofold: First, we
use this formalism to compute the spectral function of ${}^{4}$He in
anticipation of the recently measured, but not yet fully analyzed,
$A1$-collaboration data from
Mainz~\cite{Flor98,Flor01,koz00,koz01a,koz01b}. Second, we take
advantage of the L/T separation at Mainz to introduce what we regard
 as the cleanest physical observable from which to extract the nucleon
 spectral function.

\section{Formalism}

We provide here a brief discussion of our formalism. We use a
plane-wave formalism and incorporate no distortions. Our rationale for
this is that we concentrate on polarization observable which are
typically insensitive to distortions. Moreover, in some occasions the effect
of distortions is determined from other treatments and thus we
are able to concentrate on the fundamental physics with no diversions. Notably, there is a definite appeal in terms
of practicality: we can use now the  Gardner's and
Piekarewicz's~\cite{gp94} trick which renders transparent analytical results
for all observables.

The  Gardner and Piekarewicz trick enables us to introduce the concept
of a ``bound-state propagator'':
\begin{eqnarray}
  S_{\alpha}({\bf p})  
  &=& {1 \over 2j+1} \sum_{m} 
               {\cal U}_{\alpha,m}({\bf p}) \,
      \overline{\cal U}_{\alpha,m}({\bf p}) \nonumber \\
  &=& ({\rlap/{p}}_{\alpha} + M_{\alpha}) \;, \quad
      \Big(\alpha=\{E,\kappa\}\Big) \;.
\end{eqnarray}
The mass-, energy-, and momentum-like
quantities in this expression are defined in terms of the upper component of the Dirac spinor
$g_{\alpha}(p)$ and the
lower component of the Dirac spinor $f_{\alpha}(p)$~\cite{abpi2000,thesis,abpi2001}.

The evident similarity in structure between the free and bound
propagators for the direct product of spinors results in an
enormous simplification; we can now employ the powerful trace
techniques developed by Feynman to evaluate all
observables --- irrespective if the nucleon is free or bound to a
nucleus.  It is important to note, however, that this enormous
simplification would have been lost if distortion effects would have
been incorporated.

In order to automate the straightforward but
lengthy procedure of calculating these Feynman traces, we rely on the {\it FeynCalc 1.0}\cite{mh92}
package with {\it Mathematica 2.0} to calculate all traces involving
$\gamma$-matrices.

\section{Results}
\subsection{kaon quasifree process}
\begin{figure}[h]
  \epsfig{figure=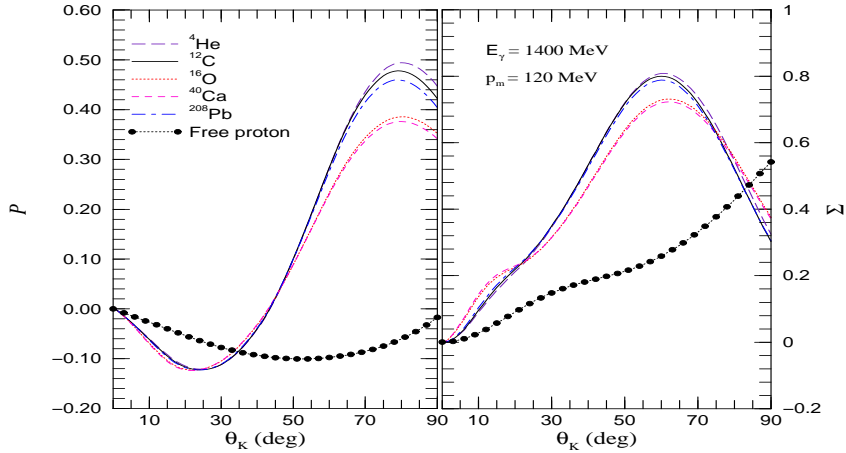,height=4.5in,width=2.45in,angle=-90}
 \caption{The polarization observables for the knockout of a valence
  proton from a variety of nuclei and for a free proton. }
\label{kfig1}
\end{figure}

We start the discussion of our results by examining the nuclear
dependence of the polarization observables. Fig.~\ref{kfig1} displays
%%%%%%
%%%
%%%
the recoil polarization (${\cal P}$) of the $\Lambda-$hyperon and the
photon asymmetry ($\Sigma$) as a function of the kaon scattering angle
for the knockout of a valence proton for a variety of nuclei, ranging
from ${}^{4}$He all the way to ${}^{208}$Pb. These observables were evaluated at a photon
energy of $E_{\gamma}\!=\!1400$~MeV and at a missing momentum of
$p_m\!=\!120$~MeV. We  have used the Saclay-Lyon model for the elementary
amplitude~\cite{dfls96}. We have included also
polarization observables from a single proton to establish a baseline
for comparison against our bound--nucleon calculations. The
sensitivity of the polarization observables to the nuclear target is
rather small.  Moreover, the deviations from the free value are significant. This indicates important
modifications to the elementary process in the nuclear
medium. Although not shown, we have studied the importance of relativity and
found that these observables are insensitive to relativistic dynamics.

\begin{figure}[h]
\begin{center}
  \epsfig{figure=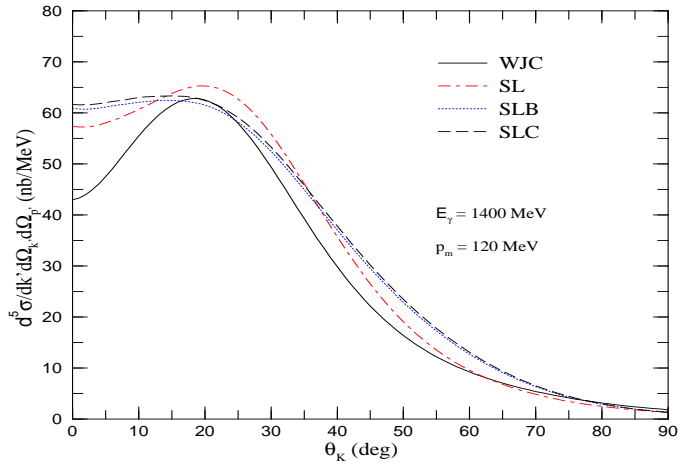,height=3.5in,width=2.45in,angle=-90}
 \caption{The differential cross section for the knockout of a proton
	  from ${}^{12}$C using various models for the elementary
	  amplitude.}
\label{kfig2}
\end{center}
\end{figure}

\begin{figure}[h]
  \epsfig{figure=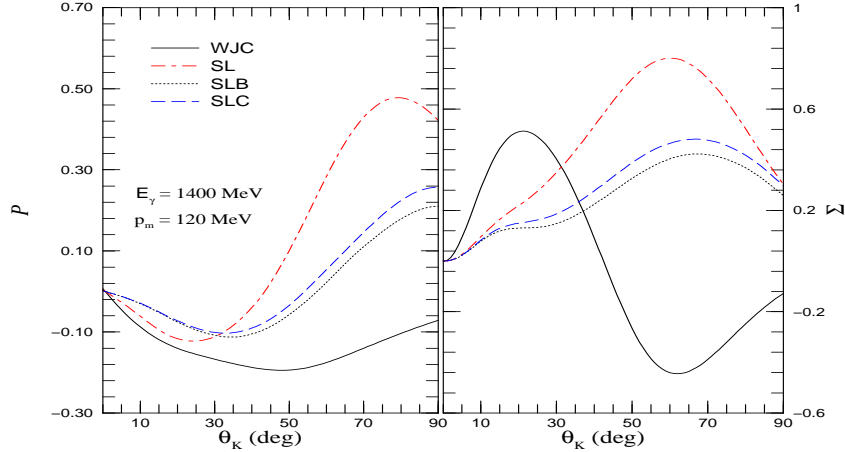,height=4.5in,width=2.45in,angle=-90}
 \caption{The polarization observables for the knockout of a proton
	  from ${}^{12}$C using various models for the elementary
	  amplitude.}
\label{kfig3}
\end{figure}

Having established the independence of polarization observables to
relativistic effects and to a large extent to the nuclear target we
are now in a good position to discuss the sensitivity of these
observables to the elementary amplitude (note that an insensitivity of
polarization observables to final-state interaction has been shown in
Ref.~\cite{blmw98}). We display in Fig.~\ref{kfig2} the differential
%%%
%%%
cross section as a function of the kaon scattering angle for the
knockout of a proton from the $p^{3/2}$ orbital in ${}^{12}$C using
four different models for the elementary
amplitude~\cite{dfls96,wcc90,mfls98}. Although there are noticeable
differences between the models, primarily at small angles, these
differences are relatively small. Much more significant,
however, are the differences between the various sets for the case of
the polarization observables displayed in Fig.~\ref{kfig3}. The added
%%%
%%%
sensitivity to the choice of amplitude exhibited by the polarization
observables should not come as a surprise; unraveling subtle details
about the dynamics is the hallmark of polarization observables. In
particular, polarization observables show a strong sensitivity to
the inclusion of the off-shell treatment for the various high-spin 
resonances, as suggested in Ref.~\cite{mfls98}

\subsection{electron quasifree process}
There is a vast amount of literature on $(e,e'p)$ reaction in the 
quasifree region. Most relevant to our present discussion is the one 
pertaining to fully  relativistic 
calculations such as the extensive set of studies conducted by the 
{\it ``Spanish''} group of Udias and 
collaborators~\cite{Ud93,Ud96,Ud97,cdmu98,Ud98,Ud99,Ud01}. 
These studies have shown that the many subtleties intrinsic to the
relativistic approach challenge much of the ``conventional wisdom''
developed within the non-relativistic framework and that, as a result,
a radical revision of ideas may be required.

The experimental extraction of the spectral function 
is based on a non-relativistic plane-wave result that is typically
referred to as the factorization~\cite{Fru84}:
%%%
\begin{equation}
S(E , {\bf p}) = {1\over p^\prime E_{p}^{\prime}\sigma_{eN}} \;
\frac{d^6\sigma}  
  {d{E^{\prime}_e d\Omega_{{\bf k}^{\prime}}
  {d{E}^{\prime}_p d\Omega_{{\bf p}^{\prime}}}}} \;.
 \label{speca} 
\end{equation}
However, this procedure is problematic. First, the quasifree cross
section [the numerator in Eq.~(\ref{speca})] suffers from the
off-shell ambiguity; different on-shell equivalent forms for the 
single-nucleon current yield different results. Second, the problem 
gets compounded by the use of an elementary electron-proton cross 
section ($\sigma_{eN}$) evaluated at off-shell
kinematics~\cite{forest83}. Finally, the projection of the bound-state 
wave-function into the negative-energy sector as well as other
relativistic effects spoil this assumed
cross section factorization~\cite{cdmu98}.      

To be noted here that the projection of the bound-state spinor 
into the negative-energy states dominate at large missing momenta 
and may mimic effects perceived as ``exotic'' from the non-relativistic 
point of view, such as an asymmetry in the missing-momentum 
distribution~\cite{gp94} or short-range correlations~\cite{pr92}. 
Indeed, Caballero and collaborators have confirmed that these 
contributions can have significant effect on various observables, 
especially at large missing momenta~\cite{cdmu98}.

While a consistent relativistic treatment seems to have spoiled the
factorization picture obtained from a non-relativistic analysis, and
with it the simple relation between the cross-section ratio and the 
spectral function [Eq.~(\ref{speca})], the situation is not without
remedy. Having evaluated all matrix elements analytically in the plane-wave limit, the source of the
problem can be readily identified in the form of several ambiguous
kinematical factors when evaluated off-shell. Thus we search for an observable that
exhibits a weak dependence on these quantities and we find, perhaps not 
surprisingly, that the longitudinal component of the hadronic tensor 
could be such an observable which is given (in
parallel kinematics) by
%%%
\begin{eqnarray}
 R_{\rm L} \equiv W^{00}
  \simeq F_1^2 (E_{p}^\prime+M) \;\rho(p), 
\end{eqnarray}
%%%
where $\rho(p)$ is nothing but the momentum distribution of the bound nucleon. 

This expression depends on unambiguous kinematics quantities and is valid
up to small (1-3 \%) second-order corrections. It is also independent of the small components 
of the Dirac spinors and of the negative-energy states. 
Moreover, it is free from of off-shell ambiguities.

\begin{figure}[h]
  \epsfig{figure=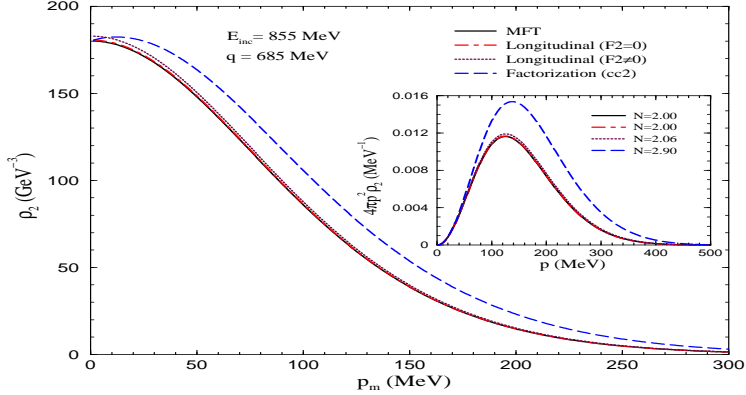,height=4.5in,width=2.45in,angle=-90}
 \caption{The proton momentum distribution $\rho_2$ for ${}^{4}$He as
	  a function of the missing momentum extracted using various methods. The inset shows the
	  corresponding integrand from which the shell occupancy may
	  be extracted.}
\label{efig1}
\end{figure}

The momentum distribution for ${}^{4}$He is displayed in
Fig.~\ref{efig1} using various methods for its extraction.  The solid
%%%
%%%
line gives the ``canonical'' momentum distribution, obtained from the
Fourier transform of the $1S^{1/2}$ proton wave-function. The momentum distribution extracted from the
longitudinal response (dot-dashed
line) is practically indistinguishable from the canonical momentum
distribution.  To be noted here that the contribution from the anomalous form factor $F_{2}$ to
the longitudinal response (see the dotted line in the figure) is small because it appears 
multiplied by two out of three ``small'' quantities in the
problem. 

The last calculation displayed in Fig.~\ref{efig1} corresponds
to a momentum distribution extracted from the factorization 
approximation (long dashed line). The momentum distribution extracted 
in this manner overestimates the canonical momentum distribution 
over the whole range of missing momenta and integrates to 2.9 
rather than 2; this represents a discrepancy of 45 percent.

\begin{figure}[h]
  \epsfig{figure=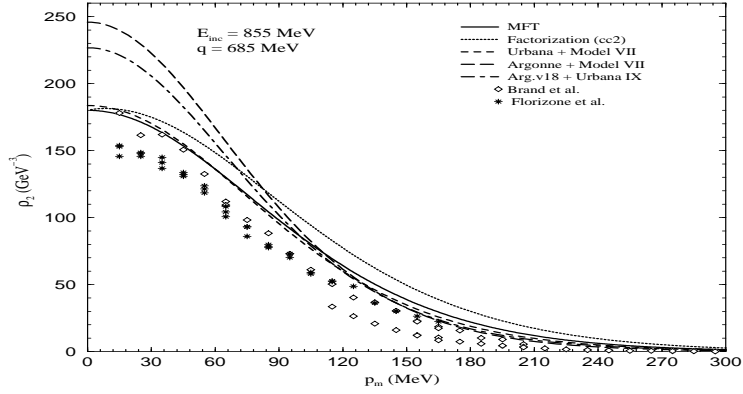,height=4.5in,width=2.45in,angle=-90}
 \caption{A comparison between our relativistic calculations, 
	  non-relativistic calculations reported elsewhere,
          and experimental data for the proton momentum 
	  distribution in ${}^{4}$He.}
\label{efig2}
\end{figure}

In Fig.~\ref{efig2} a comparison is made between our results and 
%%%
%%%%
non-relativistic state-of-the-art calculations of the momentum
distribution of ${}^{4}$He. The solid line displays, exactly 
as in Fig.~\ref{efig1}, the canonical momentum distribution. 
We see no need to include the momentum distribution extracted 
from the longitudinal response as it 
has been shown to give identical results.

In addition to our own calculation, we have also included the
variational results of Schiavilla and collaborators~\cite{sch86}, for
both the Urbana~\cite{Lag81} (dashed line) and the
Argonne~\cite{Wir84} (long-dashed line) potentials, with both of them
using Model VII for the three-nucleon interaction. The variational
calculation of Wiringa and collaborators~\cite{Wir91,Wir01,forest96}
(dashed-dotted) has also been included; this uses the Argonne v18
potential~\cite{Wir95} supplemented with the Urbana IX three-nucleon
interaction~\cite{Pub95}. Figure~\ref{efig2} also shows NIKHEF data by
van den Brand and collaborators~\cite{brand91,brand88} as well as
preliminary data from MAINZ by Florizone and
collaborators~\cite{Flor98,Flor01}.
Comparisons to the preliminary Mainz data of Kozlov and
collaborators~\cite{koz00,koz01a,koz01b} have also been made (although
the data is not shown). Thus, high-quality data for the
momentum distribution of ${}^{4}$He is now available up to a missing
momentum of about 200~MeV. We find the results of Fig.~\ref{efig2}
quite remarkable. It appears that a simple relativistic mean-field
calculation of the momentum distribution rivals --- and in some cases
surpasses --- some of the most sophisticated non-relativistic
predictions. 
Still, theoretical predictions of the momentum distribution
overestimate the experimental data by up to 50-60\%. Part of the
discrepancy is attributed to distortion effects which are estimated
at about 12\%~\cite{Flor98,sch90}. However, distortions are not able
to account for the full discrepancy. We have argued earlier that an
additional source of error may arise from the factorization
approximation used to extract the spectral
function from the experimental cross section. We are confident that the approach
suggested here, based on the extraction of the spectral function from
the longitudinal response, is robust. While the method adds further
experimental demands, as a Rosenbluth separation of the cross section
is now required, the extracted spectral function appears to be weakly
dependent on off-shell extrapolations and relativistic effects.

\section{Conclusions}
\label{sec:concl}

We have developed a relativistic formalism for studying quasi-free
processes from nuclei. The formalism can be applied with ease to a
variety of processes and renders transparent analytical expressions
for all observables. We have applied it to the processes of kaon
photoproduction and electron scattering.

For the kaon quasifree process, we have found that the polarization
observables are very sensitive to the fundamental physics in this
process, but at the same time mostly insensitive to distortion
effects, relativistic effects, and nuclear target effects. We conclude
that the polarization observables are one of the cleanest tools for
probing both the elementary amplitude ($\gamma p \rightarrow K^+
\Lambda$) and nuclear medium modifications.

For the electron quasifree process, we have derived a robust procedure
for extracting the momentum distribution using the longitudinal response.
Furthermore, we found that the relativistic mean-field calculation of the momentum 
distribution in ${}^{4}$He rivals --- and in some cases surpasses --- 
some of the most sophisticated non-relativistic predictions to date. 

\section*{Acknowledgments}
  This work was supported 
in part by the United States Department of Energy under Contract
No. DE-FG05-92ER40750 and in part by a joint fellowship from the 
Japan Society for the Promotion of Science and the United States 
National Science Foundation.

\end{document}